# Proximity effects in Graphene and Ferromagnetic CrBr$_3$ van der Waals Heterostructure

Sushant Kumar Behera[†], Mayuri Bora[†], Sapta Sindhu Paul Chowdhury and Pritam Deb*

Department of Physics, Tezpur University (Central University), Tezpur-784028, India

*Corresponding Author: pdeb@tezu.ernet.in
[†]Authors have equal contribution

**Abstract**

We report on first-principle calculations on magnetic proximity effect in a van der Waals heterostructure formed by a graphene monolayer induced by its interaction with a two-dimensional (2D) ferromagnet (chromium tribromide, CrBr$_3$). We observe that the magnetic proximity effect arising from the spin-dependent interlayer coupling depends sensitively on the interlayer electronic configuration. The proximity effect results in spin polarization of graphene orbital by up to 63.6 %, together with a miniband splitting band gap of about 73.4 meV and 8% enhancement in magnetic moment (3.47μ$_B$/cell) in heterostructure. The position of the Fermi level in the Dirac cone is shown to depend strongly on the graphene-CrBr$_3$ interlayer separation of 3.77Å. Consequently, we also show that a perpendicular electric field can be used to control the miniband spin splitting and transmission spectrum. Also, the interfacial polarization effect due to the existence of two different constituents reinforces the conductivity via electrostatic screening in the heterolayer. These findings point toward potential nanoscale devices where the electric field driven magnetic proximity effect can lead to unique spin controllability and possible engineering of spin gating.

## 1 Introduction

Magnetic proximity effect (MPE) is one of the most essential interfacial phenomena giving an idea to design two dimensional (2D) magnetic heterostructure [1]. Generally, two materials of different lattice arrangements couple together when they are brought into close vicinity due to proximity effect [2,3]. The impact of proximity effect results in the improvement of their properties and leading to variation of interfacial exchange coupling across the interface [4]. Recently, MPE has been investigated via exchange coupling between non-magnet and magnetic substrates [5], non-magnetic bilayer systems [6]. In these cases, it is a challenging task to establish long-range magnetic ordering [7] and the magnetic moment induced from the substrate via proximity effect in non-magnetic material has limited effect in charge and spin-based transport properties [8]. Apart from the magnetic proximity effect, van der Waals (vdW) heterostructures are quite flexible with incorporation of extrinsic perturbation like strain [9], electric-field [10, 11] and magnetic-field [12] for its feasibility in device engineering. Graphene based heterostructures have been explored in graphene-hBN based transistor application [13] via effective gate tuning [14] and ballistic transport in ambient condition [15]. The electronic structure properties can be controlled via external effects in vdW heterostructure system [16]. Robust exchange interaction has been demonstrated theoretically where valley splitting is observed upto 70 meV in graphene-Bismuth Ferrite (BFO) heterostructure [17]. Hybrid heterostructures like platinum- yttrium iron garnet (Pt/YIG) [18], graphene-yttrium iron garnet (Gr/YIG) [19] have been studied for realizing anomalous Hall Effect and topological phases.

2D graphene is considered as one of the model systems for nanospintronics applications for its suitability in high electron mobility and long spin relaxation length [20]. Moreover, spin in graphene can be tuned via local magnetic ordering, electric field and strain to obtain new material behaviour. Thus, the idea of spin injection into graphene by proximity effect is an interesting and developing field of research in the current time. However, a fundamental challenge lies in the development of external ways to tune (i.e. gating) the transmission of spin (i.e. spin-polarization) currents at room temperature, in view of designing spin logics devices. It is realized that the growth of graphene on magnetic metallic substrates has been reported as a route to tailor graphene spin properties [21]. As a result, magnetic conducting substrates fundamentally restrict the design of novel types of spin switches via natural short-circuiting the graphene layer. Meanwhile, 2D magnetism and low-power spintronic devices are current research opportunities due to the recent discovery of long-range magnetic order in atomic thin crystals [22]. As a result, the 2D layered magnets showcase a significant tunability in the interlayer magnetic order via external electric or magnetic fields [23, 24]. It is observed that 2D layered magnetic materials can provide an efficient platform for spin control in the nonmagnetic layer via the magnetic proximity effect when integrated with other 2D materials forming van der Waals (vdW) heterostructures [25]. This implies new opportunities to realize via introducing magnetic insulator to induce magnetism in graphene by the proximity effect in the designed vdW heterostructures.

However, its possibilities are limited in spintronics due to the absence of magnetic ordering and its weak spin-orbit coupling. The limitation in graphene has triggered tremendous efforts in extrinsically induce local magnetic moments via various methods like doping or introducing magnetic material [26], induce defect engineering [27], or by introducing a magnetic substrate on which graphene layer is stacked [28]. Graphene layer with magnetic substrate is promising but, it has limitations in establishing strong magnetic ordering and intrinsic moment induced from the substrate [7]. These limitations have enabled exploring possibilities on other 2D crystals such as Chromium trihalides ($CrBr_3$) with intrinsic magnetic moment and a band gap ranges (1-2.6 eV) [9, 29, 30]. The 2D ferromagnetic semiconductor combined with graphene to form magnetic heterostructure will be superior in proximity as well as electric field modulations which are generally used as perturbative effect to vary the electronic property of the 2D heterolayer system. Van der Waals heterostructures like graphene-EuS [31], $MoS_2$-EuS [32], graphene coupled with $CrI_3$ [33], $WS_2$-$MnO_2$ heterostructure [34] has been studied for spin-valves and valley-splitting.

The motivation behind the present work emerged from the successful experimental realization on Gr-$CrBr_3$ heterostructure [35]. In this present work, we investigate the MPE in a lattice-matched vdW heterostructure formed by a ferromagnet monolayer (i.e. $CrBr_3$) and a semimetal monolayer (i.e. graphene). This report helps in triggering the idea to explore the external perturbative effects on the functionalities of Gr-$CrBr_3$ heterostructure. It is observed that mini band splitting and spin polarization occurs in electronic structure of graphene via magnetic proximity effect induced by keeping ferromagnetic $CrBr_3$ in locus of 3.77Å. In the present work, we use an *ab initio* density functional theory (DFT) simulation on Gr-$CrBr_3$ heterostructure. Our results reveal that the magnetic moment remains invariant in monolayer $CrBr_3$ system, but a notable enhancement of 8% is observed in the hetero-bilayer system. Also, interfacial polarization effect is found to be predominant in Gr-$CrBr_3$ heterostructure. The electric-field tuning of mini band splitting and ballistic transmission is observed in Gr-$CrBr_3$ heterolayer using magnetic proximity effect which makes the heterostructure system suitable for fabricating future nanoscale devices.

**2 Computational Details**

The first principle based density functional theory (DFT) calculations have been followed employing QUANTUM ESPRESSO [36] software package which uses plane wave based basis set for computations. The electron core interactions in the graphene-$CrBr_3$ heterostructure are addressed using projector augmented wave (PAW) method for the pseudopotentials [37].

Perdew-Burke-Ernzerhof (PBE) framework of the generalized gradient approximation (GGA) [38] has been used to estimate the exchange correlation (XC) energy of the said heterostructure. For taking into account the weak van der Waals (vdW) interactions between the graphene and $CrBr_3$ layers, dispersion corrections have been used by employing DFT-D2 method as proposed by Grimme [39] with vdWs radius of 2.744 Å, 2.952 Å and 3.305 Å for carbon(C), chromium (Cr) and bromine (Br) respectively. The Hellmann-Feynmann forces that act on the ions have been minimized to 0.01 eV/Å in the system to relax the structure and to make it suitable for further density functional calculations. The interplanar distance between graphene and $CrBr_3$ sheets is kept fixed at 3.77 Å with a strain of 1.8 % with 12 Å of vacuum to control the supercell image interaction. The Kohn-Sham orbitals are expanded using the plane wave basis set for which 680 eV of cut-off energy is used. Within the scheme of Monkhorst and Pack for the integrations in the Brillouin zone, a *k*-point grid of 9x9x1 has been followed in the self-consistent calculations [40]. Total energy of the system has been calculated with an accuracy of $10^{-8}$ eV by following the Davison method for iterative diagonalization of density matrices. For non-self-consistent calculations, a much denser *k*-point grid of 27x27x1 has been used for better accuracy in the results. A path consisting of highly symmetric k-points M- Γ-K-M has been used in the band structure calculations. An external electric field has been applied perpendicular to the heterostructure in -Z and +Z directions to determine the electric field effect on band topology (refer to Fig. 3(c)). In the simulation cell, a saw-like potential simulating an electric field is added to the bare ionic potential as implemented in the steps of the simulation methodology.

To perform complex band structure and transmission coefficient calculations, we have considered the scattering formulism of PWCOND [41, 42] modules of QUANTUM Espresso codes which is based on *ab initio* DFT framework. Here, the quantum-mechanical scattering channel is fixed and simulated to find the reflection ($r_{ij}$) and transmission ($t_{ij}$) amplitudes for electron waves (*j*) propagating towards the right direction from the left. Meanwhile, the total transmission at the Fermi energy is given as, $T(E_F)=\sum_{ij}|t_{ij}(E_F)|^2$, which is used later to obtain the linear ballistic conductance (*G*) via Landauer-Buttiker formula, $G=(e^2/h)T(E_F)$ implementing non-equilibrium Green's function (NEGF) in the calculation. The transmission spectra and associated parameters such as the transmission eigenstates and the transmission pathways are simulated in this work using the self-consistent NEGF method. Fast yield capacity and acceptability of accurate simulations for supercell of vdW heterostructure systems are primary reasons of implementing the process for transmission calculations [43]. Further details about the calculation are provided in Supplementary Information.

## 3 Results and Discussions

The geometry and corresponding interlayer distances are obtained by employing the surface atomic structure optimization via Feynman-Hellman theorem (shown in Supplementary Information Table S1) [44]. Computationally, this is realised by the Broyden-Fletcher-Goldfarb-Shanno (BFGS) algorithm. The hetero-bilayer has minimum energy configuration when ferromagnetic ordering is considered. Therefore, structural optimization has been done considering ferromagnetic ground state (shown in Supplementary Information Fig. S1 and S2). The atomic configurations are shown in Figure 1. The interlayer distance between graphene and $CrBr_3$ (Gr-$CrBr_3$) is 3.77 Å. The interlayer interaction is considered as van der Waals type. Total optimized energy is simulated as a function of interplanar distance (shown in Fig. 1(b)) to fix the optimal interplanar spacing for close proximity coupling between graphene and $CrBr_3$. The electronic band structure of Gr-$CrBr_3$ heterostructure is plotted at zero biasing, shown in Fig. 1(d).

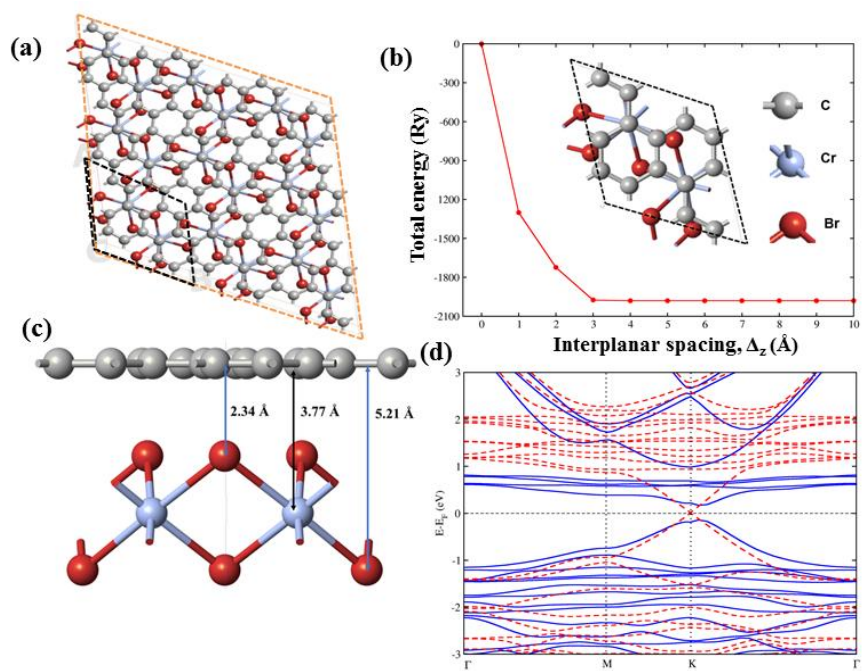

**Figure 1:** Crystal structure of Gr-$CrBr_3$ heterolayer. (a) top view of Gr-$CrBr_3$ supercell (b) Variation of total energy as a function of interplanar spacing [*inset* shows top view of Gr-$CrBr_3$ unit cell] (c) side view of Gr-$CrBr_3$ unit cell. The grey, blue and red balls describe C, Cr and Br atoms, respectively. (d) Electronic band structure of the heterostructure at absence

of external electric field. Blue continuous lines and red dotted lines show the up and down spin states, respectively.

The Partial density of states provides a description regarding the contribution of atomic orbitals individually in Gr-CrBr$_3$ heterostructure. Fig. 2 shows the spin-polarized PDOS pattern of the Gr-CrBr$_3$ heterostructure. In Fig. 2(a), it is seen that the state of π-bond delocalization and strong hybridization of p-orbitals of carbon atoms are present in heterostructure whereas conduction band is dominated by C-p orbitals with weakly hybridized s-orbitals in both spin-up and spin-down direction. Fig. 2(b) indicates that the conduction band is influenced by Cr-4d$^2_z$ orbital having intense peak for both spin-up and spin-down states. Hybridization of Cr-d orbitals is more in Gr-CrBr$_3$ heterostructure. Fig. 2(c) illustrates the maximum contribution of Br-p orbitals in the conduction band and intense peak of Br-s orbitals is observed for both spin-up and down direction in the valance band. *s*-orbitals almost have pure element-independent character exhibiting a significant spilt energy gap from -5 eV to 0.1 eV. The contribution of Br-p orbitals signifies with the intense peak in the conduction band and there is no significant role of s-orbital in the conduction band. Therefore, it can be seen that p- orbital of C and d-orbital of Cr are taking active participation in the bonding of Gr-CrBr$_3$ heterostructure. The two orbitals p- and d-get tuned when they come in close proximity. The spin-polarization effect (*P*) is explained by $P = \left| \frac{N_\uparrow(E_F) - N_\downarrow(E_F)}{N_\uparrow(E_F) + N_\downarrow(E_F)} \right|$ where, $N_\uparrow(E_F)$ and $N_\downarrow(E_F)$ are the majority and minority spin channel at the Fermi level, respectively. The percentage of spin-polarization is obtained from orbital resolved PDOS calculation for Gr-CrBr$_3$ found to be 1.5 %, 71.2% and 0.8% for carbon, chromium and bromine, respectively. It is obvious that the spin-polarization percentage for chromium will be significantly more than carbon and bromine due to the presence of inherent magnetic property in chromium. Similar trend of spin polarization is calculated for heterostructure system averaging 63.9% (shown in Fig. 2(d)) which clearly indicates the dynamic nature of all atomic orbital spins due to the proximity coupling between semimetal graphene and ferromagnetic CrBr$_3$.

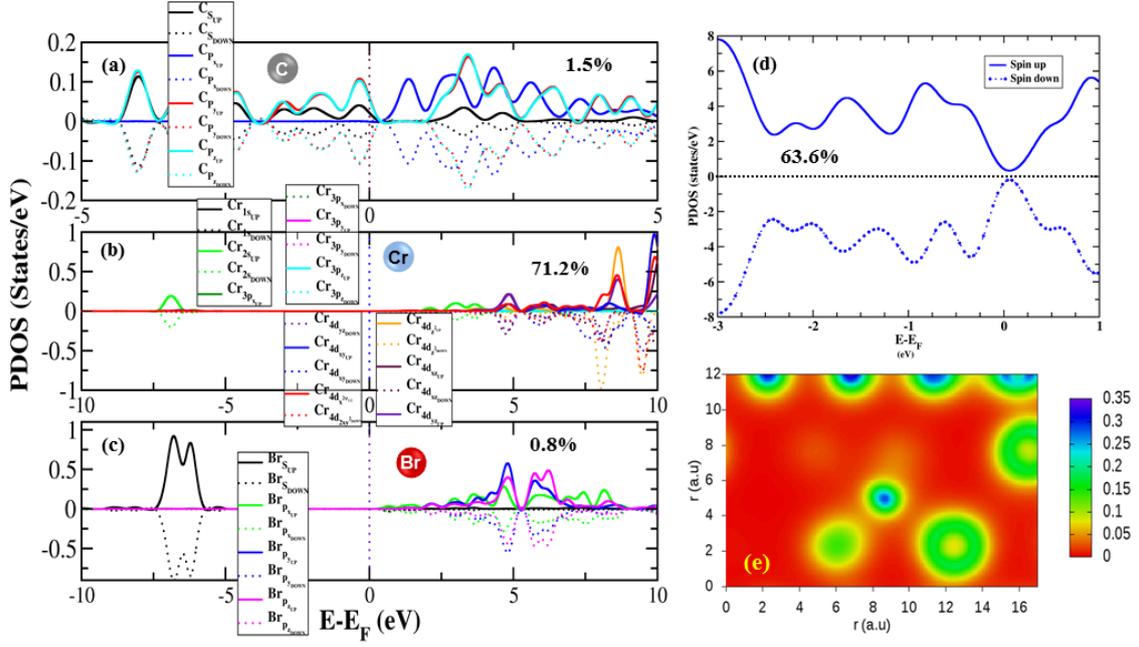

**Figure 2:** Electronic structure of heterolayer of Gr-CrBr$_3$ at zero biasing. The calculated spin polarized partial density of states (PDOS) of Gr-CrBr$_3$ heterostructure. (a) Carbon in grey coloured ball (b) Chromium in blue coloured ball and (c) Bromine in red coloured ball. (d) PDOS of Gr-CrBr$_3$ heterostructure system with polarization percentage of up and down spin states. (e) Charge density plot for Gr-CrBr$_3$ heterolayer.

Figure 2(e) describes the charge density profile of Gr-CrBr$_3$ heterostructure. As we know, that the monolayer of CrBr$_3$ has a semiconducting nature with a finite band gap while the pristine monolayer graphene is a semimetal with a zero band gap. Hence, the heterostructure system has a distinctive semimetal-semiconductor associated characteristic. It is observed that there is no adequate transfer of charge due to the vdW interaction between the two layers, but distribution of charge on the surface of Gr-CrBr$_3$ heterostructure is found. The robustness of proximity effect is mostly dependent on the orbital hybridization at the interface. The blue regions show localization of electrons in the core. Moreover, the contribution of p, d-orbitals for C and Cr also signifies from the PDOS where C$_p$ and Cr$_d$ orbitals are more intense than the Br$_p$ orbitals. This describes the delocalization of electrons identified from green and yellow region. We have also plotted the charge density contour plot where the contour line shows the electron localization and delocalization in Gr-CrBr$_3$ heterostructure is shown in supplementary information Fig. S3. The magnetic ordering in the heterostructure is found to be ferromagnetic in ground state with magnetic moment 3.47$\mu_B$/cell. Moreover, the magnetic moment of CrBr$_3$ monolayer is found to be 3.39$\mu_B$/cell with an increment of 39% compared to the bulk counterpart value of 3.00μ$_B$/cell [45] at zero external electric field. There is an increment in magnetic moment upto 8% in the heterostructure compared to monolayer CrBr$_3$ indicating the magnetic proximity effect.

Inclusion of semimetallic graphene layer with CrBr$_3$ exerts a staggered potential on CrBr$_3$ interface at a finite distance due to which the electron density and charge distribution are modulated in the heterolayer system. To have a clear realization regarding these interface polarization phenomena, the Fermi energy diagram has been demonstrated schematically in Fig. 3(a) and (b). Perpendicular electric field has been applied to the heterosystem in both positive and negative direction of Z-axis, while the magnetic ordering is imposed along X-axis following the ferromagnetic ordering. The field variation is considered in V/Å unit along the specified direction shown in the Fig. 3(c) and (d). Here, in the present case of calculations, application of the perpendicular external electric field, which causes a redistribution of charges, is not likely to lead any structural changes in the surface atoms of the heterolayer system as previously observed in silicene based hetero layers [46]. In Fig. 3(e), a prototype of single gate field-effect transistor (SG-FET) device from the Gr-CrBr$_3$ heterostructure is shown. Gr-CrBr$_3$ is sandwiched between two electrodes (source and drain) in the central region. The gate is placed above the central region, so that the flow of electrons is possible through the channel from source to drain when applying external electric field.

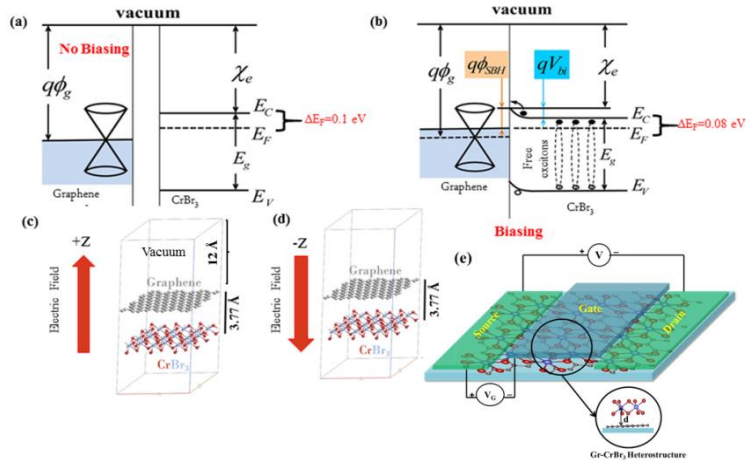

**Figure 3:** Schematic representation of Gr-CrBr$_3$ hetero-bilayer (a) without and (b) with biasing. Application of external electric field in (c) positive (+Z direction) and (d) negative (-Z direction) bias (e) the proposed device model for SG-FET device.

It is observed that there is a fluctuation in the split-off energy gap with respect to biasing voltage from 0 to ±0.5V/Å shown in Fig. 4(a). Thus, distinct nonlinearity is marked in this biasing leading to interlayer polarization. Similarly, a decreasing trend (*or* increasing trend) is observed while increasing the field in reverse (or positive) bias from 0 to -0.5 (or +0.5) V/Å, indicating attractive (or repulsive) force due to the interlayer polarization. The energy gap closing marks the tendency of variation in the direction of reverse electric field which is opposite to the in-built electric-field effect and can modulate the charge transport at the interface. Meanwhile, the mixed (*i.e.* linear and nonlinear) behaviour supports to investigate the charge transmission spectra via ballistic transport calculation to realize the proximity effect in heterostructure and role of the interlayer polarization.

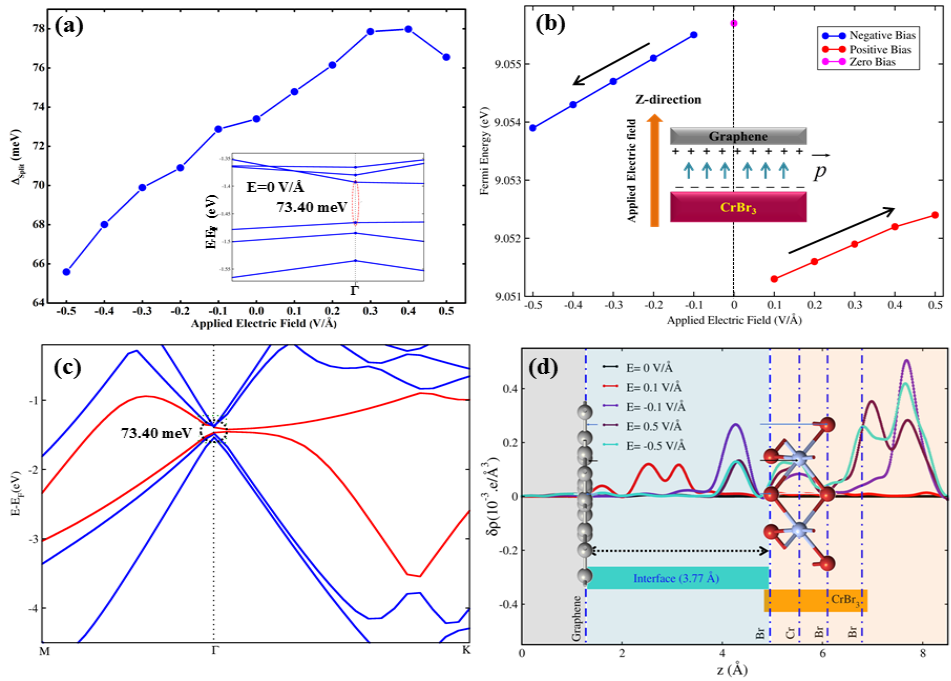

**Figure 4:** (a) The variation in split-off energy as a function of field of the heterostructure (inset shown the zoomed version of energy split-off gap 73.40 meV at 0 V/Å field). (b) Variation in Fermi energy due to the biasing via external electric field in the range -0.5 to +0.5 V/Å. The schematic of interfacial polarization is shown as inset of (b) under biasing. (c) The energy split-off band structure at 0V/Å field. (d) Induced electrostatic charge density, $\delta\rho$ for Gr-CrBr$_3$ heterostructure at different bias voltage for negative charge (-$q$) values. The atomic highlights of Gr-CrBr$_3$ interface is shown as dotted vertical blue lines to note the unbalanced formation of electric dipole moments between graphene (C atom) and CrBr$_3$ (Cr and Br atom) accordingly with the number of atomic layers used to form the heterostructure.

The modulation in Fermi energy level has been plotted against applied field in both directions to have a comparative understanding (shown in Fig. 4(b)). In the heterostructure, it is observed that non-linearity is present in both forward and reverse biasing enhancing the nonlinearity factor. As a result, the edge polarization is effective to form collective interlayer polarization. The split-off energy level of heterostructure is shown as band structure plot at 0V/Å (shown in Fig. 4(c)). Furthermore, the modulation in energy gap for the heterostructure is shown in supplementary information Fig. S4 to S13. As the magnetic moment (3.39$\mu_B$/cell) remains invariant for $CrBr_3$ monolayer with biasing, while a minor distinct fluctuation (±0.06$\mu_B$/cell) in moment value (3.47$\mu_B$/cell) is observed in case of heterostructure (shown in supplementary information Fig. S14). In case of a metal, the effect of electrostatic screening is intrinsically controlled by the spin-dependent screening in the ferromagnetic content which is quantum mechanical origin [47-49], shown in Fig. 4(d) for the semimetal-ferromagnet heterojunction interface. The applied electrostatic field is screened from the interior of Cr atom by the induced surface charges of both spin (up and down) projections. However, the induced surface charge is showing asymmetric majority and minority-spin contributions, leading to changes in the surface electronic and magnetic properties of the heterosystem. The relative amounts of these charge contributions have a quantum mechanical origin depending on the spin polarization of PDOS at Fermi energy level. In Fig. 4(d), the quantitative results of the charge distributions clearly demonstrate that the charge transfer in partial regions, especially for the several peaks and valleys, is significantly enhanced under a field of 0.5 V/Å. Reverse biasing of -0.5 V/Å also enhances the charge transfer in most orbital regions, compared with the case of 0 V/Å. Here, the spin-dependent screening in C atom of graphene layer is negligible compared to Cr and Br atomic layers. This seems to imply that electric field gating fails to modify the spin polarization in a nonmagnetic region leading to a minute fluctuation in induced magnetisation values with respect to biasing field in the heterostructure interface. Thus, the overall spin polarization to 63% in heterostructure system and minor fluctuation in moment value clearly shows the active presence of magnetic proximity coupling at the heterojunction. Furthermore, proximity coupling can be considered as an external perturbation which can play an important role in realizing interfacial ballistic transport behaviour of the present heterostructure system.

Fig. 5(a) represents the calculated transmission spectrum versus Fermi energy for the heterostructure. To understand the spin transport behaviour at the interface of the heterostructure, the lead-free transport property has been calculated via NEGF process. Significant step like transmission spectra have been observed from valence band to conduction band in the heterostructure. Broad peak near ~6 the transmission coefficient have been observed in the conduction band in energy window of (7.2-8.8 eV) which is due to the presence of strongly hybridized C-π states exhibited from graphene. As the C-π orbital is close enough to $CrBr_3$ monolayer, we get the transmission coefficient T(E)>2. Moreover, zero transmission coefficients have been noticed in certain energy windows (2.09-3.3 eV, 4.3-5.1 eV and 6.9-7.2 eV) which are due to no coupling effect with the leads. Though the energy of electrons between *p* and *d*-orbitals are same, but there might be blockade of spins due to the absence of leads. Because of that sudden drop in T(E) can be seen in few energy intervals.

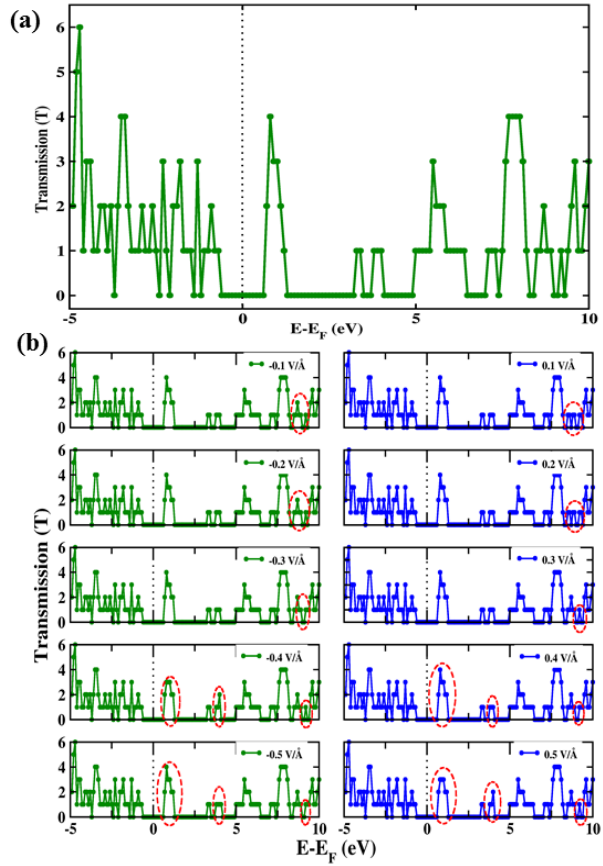

**Figure 5:** (a) The transmission coefficient calculation for Gr-$CrBr_3$ heterostructure at 0 V/Å. (b) Field induced transmission spectrum of the heterostructure at forward and reverse directions. Minor changes in the spectrum are marked using red dotted ellipses.

Transmission spectrums have been plotted with the change in electric field in forward and reverse bias (shown in Fig. 5(b)). The energy intervals observed in unbiased transmission spectrum are (0.39-2.13 eV), (3.26-4.32 eV) and (5.13-6.9 eV) which is found to be shifted and merged to the lower energy window region in the presence of electric field in forward (0.5V/Å) and reverse (-0.5V/Å) direction. It is observed that few intense and few narrow peaks are present in the spectrum which supports that the ballistic transmission is controlled through the channel with the change in electric field unlike the unbiased transmission spectrum which is suitable for field-effect transistor device application. It is observed that slight modulation occurs in the spectrum with respect to bias voltage and the transmission spectrum mostly remains unchanged. Meanwhile, the flow of ballistically generated spins can be checked from source to drain through the channel of heterostructure. The spins generated from source and detected in the drain clearly signify the switch *ON* state, otherwise if the spins are scattered away signifying the switch *OFF* state. Therefore, gating is important to trigger an effective electric field which arises due to the confinement of transport channel and electrostatic screening in heterostructure interface. This consistency nature of spectrum supports the control of interfacial polarization effect and flow continuity along the channel due to the active proximity coupling leading to its suitability in designing SG-FET device as shown in Fig. 4(e).

**4 Conclusions**

In conclusion, we investigate first-principles simulations showing that the proximity of a ferromagnetic insulator, $CrBr_3$ will induce a strong spin polarization of graphene orbitals. The $CrBr_3$ substrate is found to break the bipartite lattice of graphene into six inequivalent sublattices, resulting variable spin polarizations on the new graphene sublattices with an average spin polarization of about 63.6%. Simultaneously, a miniband split-off energy gap develops with a splitting of 73.40 meV together with a magnetic moment of 3.47μB/cell in the heterostructure, 8% larger than the monolayer $CrBr_3$ sheet. Taking into account the effect of proximity coupling at interplanar distance of 3.77 Å in forming the heterolayer, we observe the miniband state localized periodically in the interface of the heterostructure. The magnetic proximity effect in the heterojunction is reflected in the spin splitting of the miniband in electronic band structure, which provides a platform for studying the strong proximity correlation effect. An applied perpendicular electric field tunes the miniband spin splitting and transmission spectrum. Furthermore, the interfacial polarization effect activates electrostatic screening reinforcing the transmission in the heterolayer. These theoretical results deserve further experimental realization of the spin splitting effect and the field-dependent energy gap tuning in graphene-based magnetic heterostructure expecting novel nanoscale devices.

**Acknowledgements**

The authors would like to thank Tezpur University for providing High Performing Cluster Computing (HPCC) facility. The authors acknowledge to the UGC research award from University Grants Commission, Govt. of India, vide grant no. F.30-1/2014/RA-2014-16-GE-WES-5629 (SA-II). SKB (IF150325) and MB (IF180514) acknowledge to Department of Science & Technology (DST), Govt. of India for INSPIRE Fellowship. SSPC acknowledges Department of Bio-Technology (DBT), Govt. of India for financial support.

**Table of Content entry**:

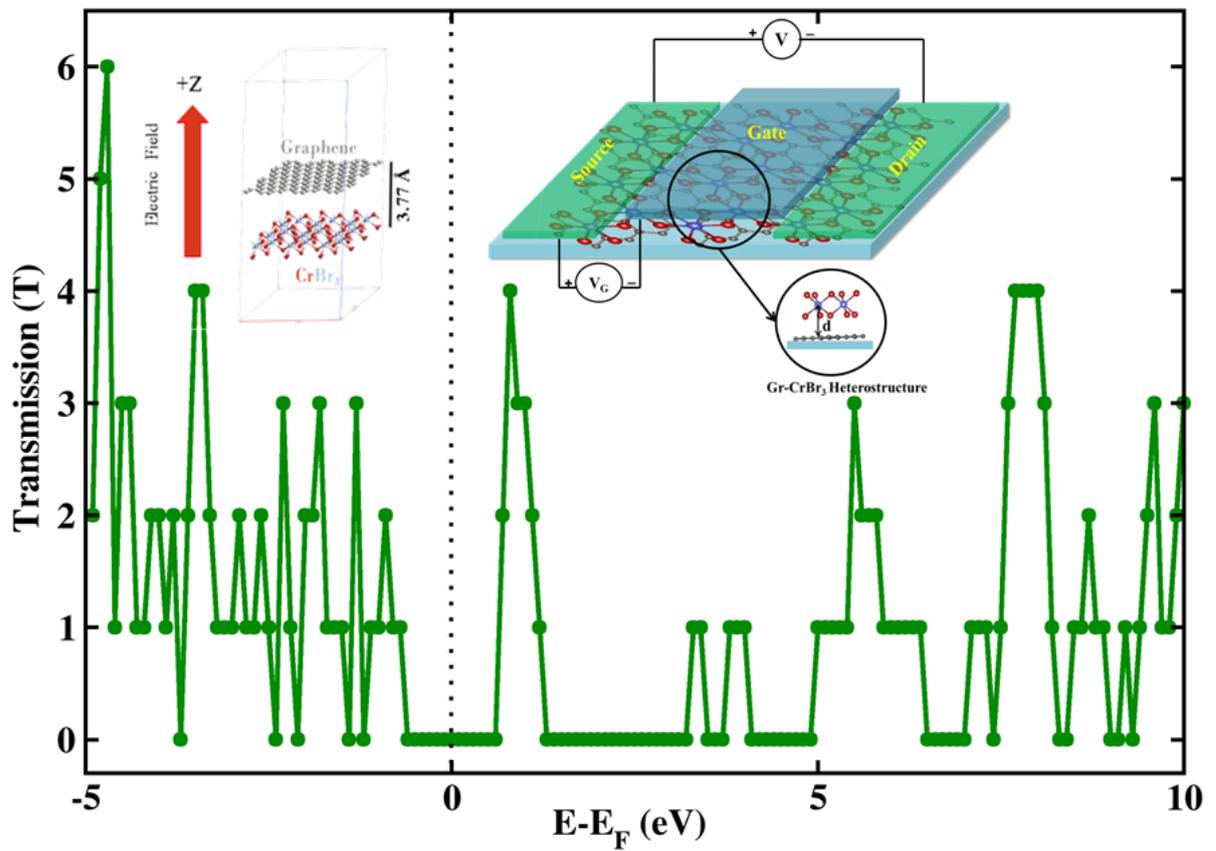

**TOC:** Schematic of magnetic proximity effect in a van der Waals heterostructure formed by a graphene monolayer induced by its interaction with a two-dimensional (2D) ferromagnet (chromium tribromide, $CrBr_3$) for designing single-gate field-effect transistor.